# The earliest drawings of datable auroras and a two-tail comet from the Syriac Chronicle of Zūqnīn

Hisashi Hayakawa (1), Yasuyuki Mitsuma (2), Yasunori Fujiwara (3), Akito Davis Kawamura (4), Ryuho Kataoka (3, 5), Yusuke Ebihara (6, 7), Shunsuke Kosaka (8), Kiyomi Iwahashi (9), Harufumi Tamazawa (4), Hiroaki Isobe (7, 10)

(1) Graduate School of Letters, Kyoto University, Kyoto, Japan
(2) Graduate School of Arts and Sciences, the University of Tokyo, Tokyo, Japan
(3) Department of Polar Science, SOKENDAI, Tachikawa, Japan
(4) Kwasan Observatory, Kyoto University, Kyoto, Japan
(5) National Institute of Polar Research, Tachikawa, Japan
(6) Research Institute for Sustainable Humanosphere, Kyoto University, Kyoto, Japan
(7) Unit of Synergetic Studies for Space, Kyoto University, Kyoto Japan
(8) Graduate School of Letters, Tohoku University, Sendai, Japan
(9) National Institute for Japanese Literature, Tachikawa, Japan
(10) Graduate School of Advanced Integrated Studies for Human Survivability, Kyoto University, Japan

**Abstract**

People have probably been watching the sky since the beginning of human history. Observers in pre-telescopic ages recorded anomalous events and these astronomical records in the historical documents provide uniquely valuable information for modern scientists. Records with drawings are particularly useful, as the verbal expressions recorded by pre-telescopic observers, who did not know the physical nature of the phenomena, are often ambiguous. However, drawings for specific datable events in the historical documents are much fewer than the verbal records. Therefore, in this paper, we show the possible earliest drawings of datable auroras and a two-tail comet in a manuscript of the Chronicle of Zūqnīn, a Syriac chronicle up to 775/776 CE to interpret their nature. They provide not only the historical facts in the realm around Amida but also information about low-latitude aurora observations due to extreme space weather events and the existence of sun-grazing comets.

**Introduction**

In this paper, we show astronomical drawings in a Syriac autograph manuscript of Chronicle of Zūqnīn to examine their nature in modern astronomy. Although modern telescopic astronomy was started by Galileo Galilei, astronomers or observers in pre-telescopic ages had recorded anomalous events in the sky that are now interpreted as supernovae (Clark & Stephenson 1977), comets (Kronk 1999), sunspots (Wittman 1978; Willis et al. 1980; Willis et al. 1996a; Stephenson & Willis 1999), or auroras (Link 1962; Keimatsu 1970; Eather 1980; Stephenson et al. 2004; Willis et al. 1996b; Willis & Stephenson 1999; Willis & Stephenson 2001; Hayakawa et al. 2015; Kawamura et al. 2016; Hayakawa et al. 2016a; Hayakawa et al. 2016b). Those records for celestial events in the historical documents frequently provide valuable information for modern scientists. Records with drawings are particularly useful, as the verbal expressions recorded by pre-telescopic observers, who did not know the physical nature of the phenomena, are often ambiguous. However, drawings for specific datable events in the historical documents are much fewer than the verbal records.

Comets have been observed since the third millennium BCE. Although there is still a controversy, the first datable comet observation was that in 673 BCE in the Babylonian records (Kronk 1999). Auroras have also been observed and recorded from an early time. The first datable aurora observation is traced back 567 BCE and found in Astronomical Diaries from Babylon (Stephenson et al. 2004), except for those appearing in the legendary records in the Bible or Chinese historical documents (Link 1962; Keimatsu 1970; Eather 1980; Silvermann 1998; Siscoe et al. 2002). Compared with the verbal records, drawings of auroras have a much shorter history. In Western Europe, the first drawing of an aurora is considered to be that on 1527/10/11 by Peter Creutzer (Eather 1980: p. 43). In China, there are earlier drawings of aurora-like phenomena, such as red vapors in *Tiānyuán Yùlì Xiángyìfù*, a manual of divination compiled under Emperor *Rénzōng* (r. 1424–1425 CE). However, these drawings are shown in the manual to explain how to interpret the celestial phenomena in the divination, and hence not related to specific datable events.

Much earlier than these drawings, it was reported by Harrak (1999) that a manuscript of the Chronicle of Zūqnīn, a Syriac chronicle compiled in the late 8th century, involves several astronomical drawings. He noted several astronomical drawings in the blanks of folios in the manuscript "Vatican Syriac 162 (hereafter Vat.Sir.162)." Although Harrak (1999, hereafter H99) did not identify the physical nature of these drawings, Neuhäuser & Neuhäuser (2015, hereafter NN15) cited two of Harrak's English translations to associate them with auroras. In this paper, we examine the drawings in question on the original manuscript of the Chronicle of Zūqnīn in comparison with related records in Syriac original texts to show these drawings, to discuss their nature, and to understand their nuances in the original texts. Our results provide vivid details for comets and intense low-latitude auroras to let scientists understand their nature further in comparison with the knowledge of modern astronomy.

**Method and Source Documentation**

In this paper, therefore, we collect and examine all drawings in the Chronicle of Zūqnīn to analyze them in the view of modern astronomy. The Chronicle of Zūqnīn is a Syriac universal chronicle spanning up to 775/776 CE. It was written by Joshua the stylite in the monastery of Zūqnīn near Amida (Diyar Bakr). Although Assemani (1728) related the author of this chronicle with Dionysius of Tellmahre, this idea was denied by Nöldeke (1896). The only manuscript is preserved in the Vatican library under the reference code "Vatican Syriac 162" (Vat.Sir.162) (Brock 1976; Brock 1979; Harrak 1999).

According to the preface of the manuscript, it was written in the Seleucid Era (SE) 1087 ("of Alexander Macedonian [*d-alleksandrāws maqedōnāyā*]" in the manuscript), corresponding to the periods between October 775 CE and September 776 CE, which we denote as 775/776 CE (f. 1v; Chabot, Chronicon II, 419)[1]. How to convert SE to CE is explained in the Appendix 1. H99 carefully examined the manuscript and concluded that the manuscript of Vat.Sir.162 is an autograph by Joshua the stylite himself. Thus, this manuscript can be regarded as written in the late 8th century by the author himself, as is described in its preface, and some of the drawings were possibly drawn after

---

[1] However, the records for 774/775 and 775/776 CE, which should have been written on the last folios of the original manuscript, are now lost.

the author's direct observation, as we discuss later.

In the case where the drawings were from the author's direct observation, the location of the observation can be specified as Amida (N 37°55′, E 40°14′), where this manuscript was written. We calculated the magnetic latitude of Amida in the late 8$^{th}$ century to be about 45 degrees based on the location of the North Geomagnetic Pole over the past 2,000 years (Merrill & McElhinny 1983). Note that the calculated value of the magnetic latitude has an uncertainty of ~5 degrees. We compared these drawings and relevant descriptions with contemporary oriental astronomical records. This approach is especially helpful in understanding the records for comets and yields further insights into observational information from the opposite end of Asia.

The source documents for this paper are as follows:
    Vat.Sir.162: The autograph manuscript of the Zūqnīn Chronicle in Vatican Library
    Chabot I: Chabot I-B (ed.) *Chronicon anonymum pseudo-Dionysianum vulgo dictum*, vol. I, Louvain, 1927.
    Chabot II: Chabot I-B (ed.) *Incerti auctoris chronicon pseudo-Dionysianum vulgo dictum*, vol. II, Louvain, 1933.

We also examined other sources to compare the descriptions. Their references are as follows:
    The Chronicle of Edessa: Guidi I (ed.) *Chronica minora* I, Paris, 1903: pp1-13. [in Syriac]
    The Chronicle of Michael the Great: Ibrahim GY (ed.) The Edessa-Aleppo Syriac Codex of the Chronicle of Michael the Great, Piscataway, NJ, 2009. [in Syriac]
    Edessene Chronicle: Wright W (ed. & tr.) The chronicle of Joshua the Stylite: Composed in Syriac A.D. 507, Cambridge, 1882. [in Syriac]
    Vita Const.: *Eusebius of Caesarea, Über das Leben des Kaisers Konstantin*, Berlin, 1991. [in Greek]
    *Shokinihongi*: Fujiwara Tsugutada (ed.) *Shokunihongi*, manuscript Ya 2-10-15 in the National Institute for Japanese Literature [in Chinese/Japanese]
    *Chénshū*: Yáo Sīlián (ed.), *Chénshū*, Beijing, 1972 [in Chinese]
    *Zhōushū*: Lìnghú Défēn (ed.), *Zhōushū*, Beijing, 1971 [in Chinese]
    *Suíshū*: Wèi Zhēng, Zhǎngsūn Wújì (ed.), *Suíshū*, Beijing, 1973 [in Chinese]
    *Nánshǐ*: Lǐ Yánshòu (ed.), *Nánshǐ*, Beijing, 1975 [in Chinese]
    *Jiùtángshū*: Liú Xù (ed.), *Jiùtángshū*, Beijing, 1975 [in Chinese]
    *Xīntángshū*: Ōu Yángxiū, Sòng Qí (ed.), *Xīntángshū*, Beijing, 1975 [in Chinese]

*Samguksagi*: Kim Pusik (ed.) (1964) *Samguksagi*, Tokyo [in Chinese/Korean]

**Result**

By examining the original manuscript of Zūqnīn, Vat.Sir.162, we found in total ten drawings of likely astronomical events: one each in folios 121v, 136v, 146r, and 155v, and two each in folios 73r, 87v, and 150v. They are summarized in Table 1. The photograph of the pages in their entirety as well as the transliteration and English translation of the corresponding parts are given in the supplementary material.

**Discussion**

**The Origin and Value of Drawings in Manuscript Vat.Sir.162**

Before discussing the physical nature of the drawings, we confirm by whom and when they were drawn, as they may be graffiti or additions made in a later period. Careful examinations of this manuscript gave us confidence that these drawings were created by the author himself, i.e., Joshua the stylite. There are three pieces of evidence that support this interpretation: the color of the inks, the typology of the handwriting in this manuscript, and mentions of the drawings in texts. Firstly, the color of the inks in the Syriac texts and those used for the drawings are similar to one another. We can recognize two inks in this manuscript: brown and red. While brown ink is used for most of the main texts, red ink is less so, used mainly for the chapter names and to highlight some remarks and specific words. The inks used for the drawings also consisted of these two colors: brown and red. Their colors look quite alike. Secondly, we found that the caption attached to Drawing 5 (f. 121v) shows very similar handwriting to that in the main texts, as was also pointed out by H99 (p. 14). Lastly and most importantly, the manuscript sometimes mentions the drawings in the text. For Drawing 6 (f. 136v), the drawing is placed in a square blank in the midst of the main text. It is very unlikely that the square blank had been left by the author of this manuscript and the drawing was added in a later period by someone else. Moreover, for Drawing 6 (f. 136v) and Drawing 10 (f. 155v), we found passages that relate Syriac texts with the drawings: "This is its shape" (f. 136v) and "it is the shape of this sign that is drawn above" (f. 155v). The above three arguments support our interpretation that the drawings are not graffiti or later additions but drawings by the author himself. Thus, these drawings have great value as contemporary sketches accompanying the texts.

Next, we discuss whether these drawings were composed after direct observation. As pointed out by H99, this chronicle has four parts, and Part IV was written on new leaves with the author's original contents, while Parts I–III were mostly copied from previous chronicles. Furthermore, a chronicle written in Edessa (the "Edessene Chronicle," now only known from the copy in Vat.Sir.162 and edited and translated by Wright (1882)) is copied between Parts II and III. This construction causes repetitions of some entries. In fact, Drawings 1 and 2 and Drawings 3 and 4 are associated with each three similar passages in folios 72v–73r and 75r–v (a part of the "Edessene Chronicle") and in f. 87v (of Part III). Chabot (Chronicon II: viii), Brock (1979) and H99 pointed out that Part III is very largely based on the chronicle of John of Ephesus.

In Part IV, the entries for each year are relatively short until 715 CE. Then, the amount of information in each entry starts to increase, and the entries from 767 CE to 775 CE are far more detailed than the older entries (H99). From this fact, we consider that the entries after 767 CE were events that were contemporary with the author Joshua the stylite and that he had much opportunity to listen to reports from contemporary direct witnesses after 716 CE. In our list, Drawings 7–10 fall into the first category, Thus, these events may have been drawn after the direct observation of the author himself, though we cannot completely discard the possibility that he drew them from hearsay. Drawing 6 (759/760 CE) is assumed to be a report based on hearsay from direct contemporary witnesses. The drawings for the older events in Parts I–III were presumably drawn from the author's imagination.

**Physical Nature of the Drawings in Manuscript Vat.Sir.162**
**Drawing 9: A Drawing for Low-Latitude Aurora "at harvest time" in 771/772 CE**
Now we discuss the physical nature of the drawings. Firstly we discuss the two strip-like drawings that were contemporary to the author, namely Drawing 9 from 771/772 CE and Drawing 10 from 773 CE. Drawing 9 corresponds to the following text:

> Another sign was seen in the northern end, its appearance testifying to the menace
> and threat of God against us. It was seen at harvest time, occupying the entire
> northern side from the eastern corner to the western corner. Its form was as follows:
> a blood-red scepter, a green one, a black one, and a saffron-colored one. It was
> going from below to above. When one scepter was extinguished, another one went
> up. And when someone was looking at it, it was changed into seventy shapes.
> (Vat.Sir.162: f. 155v; Chabot II: pp. 301–302)

This record is related with a probable aurora record by NN15 based on the English translation by H99. The northern direction is of typical low-latitude auroras. Although the observational time (i.e., during night or not) is not explicitly described (nor in Drawing 10), its observational direction, its color, and its straight structure are consistent to auroras, as we discuss below. The similarity of Drawings 9 and 10 with Drawing 4, clearly written to be observed during night, also supports this inference. Its date is not shown explicitly. Considering that it says "The sign, which was seen a year ago was also in the northern region in this year" in the relevant text of Drawing 10 (155v), it is reasonable to place this record in 1083 SE (771/772 CE), as NN15 speculated. Although NN15 also speculated its observational time i.e. "harvest time" as summer, we cannot determine its season only with this description. We must note that the Jazīra Region around Amida hosted not only summer crops, such as wheat, barley, lentil, pea, chickpea, or broad bean, but also winter crops, such as millet, cotton, rice, or sesame (Ager 2015).

Drawing 9 shows twelve double lines in a horizontal direction. However, as is shown later, another drawing in the same folio (Drawing 8) and its corresponding text indicate that the right-hand side of the folio is the upward side of the drawings. Therefore it is reasonable to consider that these stripes show a vertical structure in the sky. This interpretation is also consistent with the text: "It was going from below to above. When one scepter was extinguished, another one went up." This suggests a vertical motion as well as repeating disappearance and appearance. It reminds us of the so-called "flaming" motion of auroras (e.g., Stormer, 1955).

Its color is described as blood-red, green, black, and saffron. Red probably corresponds to the most common aurora emission from OI (630.0 nm). Green should be the OI emission in the wavelength of 557.7 nm. Black might be speculated to be "black aurora," which sometimes appears inside bright auroras. Saffron reminds us of the yellow auroras caused by OI emission in the wavelength of 557.7 nm that are seen near the horizon when the colors are influenced by the atmospheric extinction. Overall, the colorful and active appearance suggests that this aurora was similar to discrete aurora, normally observed in high latitudes caused by energetic electron precipitations. Considering that the magnetic latitude of Amida was approximately 45°, using the empirical formula suggested by Yokoyama et al. (1998), we estimated the Dst index to be at least -365 nT. This value is comparable to that for the Halloween Event in October 2003, whose value is rare and categorized as an extreme case (Tsubouchi & Omura 2007). Thus, the magnetic storms in 771/772 and 773 shown in Drawing 9 and 10 could be regarded as a large storm.

**Drawing 10: A Drawing for Low-Latitude Aurora in June, 773 CE**
Drawing 10 is similar to Drawing 9. Indeed, its corresponding text associates the event with the same kind of event seen a year earlier, most likely that depicted in Drawing 9.

> The sign which was seen a year ago in the northern region, was also seen in this year, in the month *ḥzīrān* (June), on a Friday. And equally in these three years, in which one (sign) after another was seen, each appeared on a Friday. And each equally stretched from eastern side to western side. And when someone rose up to watch it, it was changed to many different shapes in this way: when a blood-red ray was extinguished, a green one went up, and when the green one was extinguished, a saffron-colored one went up, and when this one was extinguished, a black one went up (Vat.Sir.162: f. 155v; Chabot II: pp. 301–302)

This record is in the section from 1084 SE (772/773 CE). That it happened on *'rubṭā* (Friday)[2] *ḥzirān* (June) means this event took place some Friday in June of 773 CE. Fridays in this month were on the 3rd, 10th, 17th, and 24th. This record is also ranked as a probable aurora by NN15. The drawing has seven double lines similar to those in Drawing 9. The author of this chronicle clearly related this phenomenon to that of Drawing 9, as he stated "it was seen also in this year (*etḥazyat āp b-hādē šattā*)." The descriptions of the color and motion are also very similar to those for Drawing 9, though the term of "scepter (*šabṭā*)" was replaced by the term of "ray (*zallīqā*)." In addition, the drawing itself is clearly referred to in the text as "drawn above (*ršīmā men l-'el*)," which leads us to ascertain that this is the original drawing by the author himself.

**Drawing 8: A Drawing for Heavenly Bows in May, 772 CE**
Drawing 8 is in the same folio (f. 150v) as Drawing 9 and hence is considered to be an event that took place in 1083 SE (771/772 CE), like Drawing 9. It has three "bows" drawn in red ink. The corresponding text is as follows:

> Concerning Our Lord's bow which was seen in this year and overturned, and concerning the white scepter which was seen in the middle of the sky, and which was directed toward the arc of the same bow, like an arrow. Also in this year, in the month *êyār* (May), this bow was constantly seen in the clouds, its arc being below, and its shoulders being above. And it was like a bow which was stretched by the

---

[2] It is notable that all three aurora records are placed on Friday (see, Drawings 4, 9, and 10).

> hand of a man for a battle, and also exhibiting menace and threat against the inhabitants of the earth. It was seen on a holy Sunday, at the third hour of the day… Then something like a white scepter was seen, extended from the western end of the sky, and it came to the middle of the sky, until its head turned toward the east. Then its swelling was like a rope… Then this scepter directed toward the center of the extended bow. And it showed its effect and did not delay much.

The month *êyār* (May) in 1083 SE (771/772 CE) is May of 772 CE, and the time "at the third hour (*terce*)" means 09:00 in the present time (Cabrol 1912). Since it is described as "its arc being below and its shoulders being above," the right-hand side of the folio should be the upward angle of this drawing, as already mentioned.

The physical interpretation of this record is uncertain. First of all, it was observed in the daytime and so not likely an aurora. It may be the result of a kind of atmospheric optics, such as the combination of solar pillar and parhelic arcs that resemble a bow and an arrow. It could be a typical cloud with a unique shape, a very bright comet, or a combination of them both. In any case, we do not have a convincing interpretation of the physical nature of this drawing.

**Drawing 7: A Drawing for Comet around May, 770 CE**
Drawing 7 also has the shape like a comet, drawn in the left margin of the folio:

> Concerning the sign, which was like a besom, and which was seen in the sky. In the year one thousand and eighty, in the month of *êyār* (May), one like a besom was seen, in the northeast. When it was seen in this side, it was going up and dark, as if it was collecting a dust of a house. Then at dawn it was seen, bending its lock of hair toward the earth. It was moving forward bit by bit, until it entered the axis of the wheel in the sky, where it was swallowed up and went astray. Its shape is that which is drawn above. (Vat.Sir.162: f. 146r; Chabot II: p. 257)

It is in the section of 1080 SE (768/769 CE), and the month is May (*êyār*); hence, it is dated in May 769 CE. The shape of the drawing and the text clearly indicate a comet. However, we could not find any comet records in this month in either Chinese official histories or previous studies, such as Hasegawa (1980) or Kronk (1999). There was a widely observed comet around the same month of the next year, 770 CE. For example, there are Chinese records of the comet seen from May 26 to September 25 in 770 CE (*Jiùtángshū*, *Dàizōng*: pp. 296–297; *Jiùtángshū*, Astronomy II: p. 1327; *Xīntángshū*, *Dàizōng*: p. 175; *Xīntángshū*, Astronomy II: p. 838). It was also seen in Japan near the Plow from June to August (*Shokunihongi*, XXX: f. 35a), near Aur and Tau on June 9, 770 CE and disappeared on September 9, 770 CE in Korea (*Samguksagi*, IX: f. 10b–11a), and Kronk (1999) relates all of these records to comet C770/K1.

What attracted our interest is the fact that the chronological order around Drawing 7 is in disorder, and the year goes back and forth. After this folio (f. 136v) where Drawing 6 was, it proceeds as follows: 1072 SE, 1075 SE, 1072 SE, 1076 SE, 1076 SE, 1077 SE, 1074 SE, 1078 SE, 1079 SE, 1080 SE, 1081 SE, 1080 SE (the folio of Drawing 5: f. 146r), 1078 SE, and 1084 SE (the folio of Drawing 8: f. 155v), while most of the other parts of the manuscript proceeded in a normal chronological order. From this chronological disorder, we presume there is a possibility that the year of the drawing and the associated text is actually 1082 SE (770 CE) but incorrectly written as 1080 SE (769 CE).

Actually, the motion of the comet in this text is consistent with that of C770/K1, whose orbit was calculated by Hasegawa (1979). The appearance in the north-eastern sky in the Zūqnīn Chronicle is consistent with the Chinese record of the comet appearance in Aur early in the morning on May 26, 770 CE. Also, the statement "it was going up and dark" is consistent with reports that the C770/K1 comet raised in elevation after May 26, 770 CE (Hasegawa 1979). Thus, we believe Drawing 5 is that of comet C770/K1.

**Drawing 6: A Drawing for Comet in May, 760 CE**

We now move on to Drawing 6, a comet-like one possibly drawn after hearsay from contemporaries of Joshua the Stylite himself. As one can see in f. 136v, the comet-like drawing also accompanies small circles with captions that state "Ram, Ares, Kronos (*emrā*, *arrês*, *qrāwnās*)" from left to right; hence, they are the three stars of Aries (Ram), namely α Ari, β Ari, and γ Ari, Mars (Ares), and Saturn (Kronos). It is in the section of 1071 SE (759/760 CE), and the corresponding text is following. As one can see in f. 136v, this drawing is placed in the midst of original texts and thus it can be ascertained that these were drawn simultaneously with the related texts.

> In the month *ādār* (March), a white sign was seen in the sky, before early-morning, in the north-eastern side, in the zodiacal sign which is called Aries, to the north from those three stars, which are most shining in it. And it looks like in its shape, a besom. And it was still in Aries, at its head, in the first degree (of the sign), two (degrees) from those wandering stars, Kronos and Ares, which are slightly to the south, on the twenty-second of the month. And the sign remained for fifteen nights, until the dawn of the Pentecost feast. And one end of it was narrow and duskier, one star was seen in its tip, and it was turning to the north. And the other one, being wide and darker, was turning toward the south. And it (the sign) was going bit by bit to the northeast. This is its shape:

On the dusk of the third day after Pentecost, it appeared again in the evening in the northwest. And it remained twenty and five evenings. And it was going bit by bit to the south.

And it went astray again. Then it returned and was seen in the southwest, and in this way remained there many days. (Vat.Sir.162: f. 136v; Chabot II: p. 217)

The striking part of this text is the description of two ends, one being narrow and duskier, and the other wide and darker. It may be the first description of the ion tail and the dusk tail of a comet, though the existence of two tails is not clear in the drawing. In the following, we present a deeper analysis of the text.

Firstly, we discuss its date and time. According to the text, the event was in the month *ādār* (March), so it should be March of 760 CE. The event was first seen on the 22nd of the month, remained 15 nights until the dawn of the Pentecost feast, appeared again on the third day after Pentecost, and remained another 25 days. From the text, it is also clear that the event was seen during the night. The time and the duration of the event as well as its shape, which we discuss below, are consistent with the interpretation that it was a large comet. The date of the event, on the other hand, is rather confusing. It is written that this event was seen on March 22 and lasted 15 nights up to the eve of Pentecost, but the date of the Pentecost of that year is May 25 (Grumel 1958), so it cannot be 15 nights after March 22. A probable explanation for this inconsistency is a miswriting of May (*êyār*) as March (*ādār*) in the manuscript as there is only one letter difference between them in the Syriac letter system. After a short break, this event reappeared on the eve of the third day after Pentecost and lasted another 25 nights. Thus, we conclude that this event started around May 22, 760 CE and lasted until early July of 760 CE.

This should be the record of Halley's Comet, which has an orbital period of about 75 years. According to Yeomans and Kiang (1981), Halley's Comet was at perihelion on day 20.671 (UT) of May 760 CE. The comet was also observed by Chinese astronomers. There is a record in *Jiùtángshū*, one of the official histories of the *Táng* Dynasty, stating that the comet was first observed on May 16, 760 CE within Aries and continued to be visible for about 50 days (*Jiùtángshū*, Astronomy II: p. 1324). The observed period and the association with Aries are consistent between the Chinese and the Syriac records.

As already mentioned, the outstanding feature of this record is the description of two "ends" reminiscent of the two tails of comets: the ion tail and the dust tail. The description stating that "one end of it was narrow and duskier, one star was seen in its tip and it was turning to the north" sounds like the ion tail, which is usually narrower and darker, while "And the other one, being wide and darker, was turning toward the south" sounds like the dust tail, which is usually wider and brighter. Note that "*yattīr bahhūrā*," which means "duskier," is emended to *yattīr nāhūrā* by Chabot. H99 (p. 198) accepts this emendation and interprets the appearance of the end of the sign as "more shining," contrasting with the appearance of another end that is "darker (*yattīr 'ammūṭā*)." However, we can accept the original words, if we consider that neither of the ends was dark but that their appearances were slightly different.

A preliminary simulation using the commercial software Stellar Navigator (ver.10) also shows a consistent result with the interpretations above (see the Appendix for details). On May 25, 760 CE, the relative configuration of Halley's Comet, Aries, Mars, and Saturn was the same as that in Drawing 6. Moreover, although the angle between the two tails changes with time, for most of the observed period, the dust tail extends and bends toward the south while the ion tail is relatively toward the zenith, and it increasingly turns more northward as time goes on. Thus, we conclude that Drawing 6 is of Halley's Comet, and its associated text is the oldest known description of the ion and dust tails of a comet, though the existence of the two tails is not clear in the drawing.

**Drawings 1–5: Drawings Based on Hearsays from Previous Chronicles**
The rest of the drawings are much older and hence not from the direct observation by the author. Therefore, we will describe them briefly.

Drawing 5 also has a comet-like shape. As shown in f. 121v, this drawing is accompanied by a Syriac passage, which states "Drawing of the same star (*dmūteh dīleh d-kawkbā*)." It is in the section of 885 SE (573/574 CE), and the month is *êyār* (May); hence, May of 574 CE. The associated text is provided in the supplementary material. Simultaneously, several comet observations are reported in China around May of this year. One is on May 16 in Jiànkāng (present Nánjīng) in southern China (*Chénshū*, *Xuāndì*: p. 87; *Nánshǐ*, *Chén*: p. 295). There are three separate observations at Chángān (present Xīān) in northern China: from April 4 until July 5, from May 31 until June 9, and on June 9 (*Suíshū*, Astronomy II: p. 607; *Zhōushū*, *Wǔdì*, V: p. 84). Kronk (1999) categorized these records into two comets and associated the one from April 4 (until July 5) with C574/G1 and the others to another circumpolar. According to the associated text, the comet of Drawing 5 was visible for "two or three months" and finally reached the "southern region" of the sky. Considering the duration and the motion, it is more likely that this comet was C574/G1, not the circumpolar.

Drawing 4 in f. 87v has a stripe shape like Drawings 9 and 10, with its parallel text in the Edessene Chronicle (f. 75r–v; Wright 1882: pp. 43–44 [critical edition], pp. 36–37 [translation]), the Chronicle of Edessa, and the Chronicle of Michael the Great (Ibrahim 2009, p260), as well. It is in the section of 813 SE (501/502 CE) with its date described as Friday, August 22. Therefore, we can determine its date to be August 22, 502 CE. As is already mentioned, the record of f. 87v is copied and summarized from previous chronicles. The Edessene Chronicle (original text: f. 75r–v; Wright 1882, pp. 43–44; translation: pp. 36–37), one of parallel records for this one, has much similar record as well. This record is surely of an aurora, as it was observed at night to the north and simultaneously observed not only in Edessa, where the Edessene Chronicle was compiled, but also in Ptolemais, Akko, Tyre, and Sidon, according to the Edessene Chronicle. Wright (1882: p. 37) simply indicates that it was an aurora.

Finally, we discuss Drawings 1 and 2 in f. 73r and Drawing 3 in 87v. They are in the sections of 811 SE (499/500 CE). Drawing 1 looks like three upside-down bows in the sky, while Drawings 2 and 3 look like comets. The text associated with Drawings 1 and 2 is as follows. Clearly, from the morphological description, the first paragraph is for Drawing 1 and the second for Drawing 2.

> And again in this year, in *tešrī ḥrāy* (November), three signs were seen in the sky, at midday, (each of) them resembled a bow that is seen in clouds. Their arcs were equally below and their ends were above. One of them was in the southern side, in the middle of the sky, and another in its east. And another in the west. And thus they were turned toward the sky.
>
> And again in this year, in *kānūn* (January), another sign, which resembled a lance, was seen in the sky, in the south-western side. Some people called it the besom of perdition. And others called it the lance of war. But their story is also written down by us. (Vat.Sir.162: f. 87v; Chabot II: p. 4)

There is also an almost identical text on folios ff. 72v–73r, as well, which is included in the supplementary material.

Regarding Drawing 1, its month is *tīšrīn ḥrāyā* (November), and hence it was in November of 499 CE. Its shape is rather puzzling, but since it was seen at midday, one likely interpretation is a kind of atmospheric optical phenomenon, such as parhelia, as suggested by Wright (1882), or cloud iridescence.

Drawing 2 has a comet-like shape. Drawing 3 is also related to an identical text on f. 87v, as stated above, and hence is regarded as another drawing of the same comet. It was seen in the month *kānūn trayyānā* (January), hence in January of 500 CE. As mentioned earlier, these drawings are not from the author's direct observation but imaginings. Kronk (1999) introduced this comet based solely on the Edessene Chronicle (72v-73r, Wright 1882: p32 [edited text]; p27 [translation]). Confusingly, Kronk (1999) calls this Edessene Chronicle the "Chronicle of Joshua the Stylite" (Cf. Brock 1979; Harrak 1999).

**Conclusion**

We have examined ten drawings in the autograph manuscript Vat.Sir.162 of the Zūqnīn Chronicle compiled in the late 8$^{th}$ century. Within these ones, Drawings 2, 3, 5, 6, and 7 are comets, and 4, 9, and 10 are likely auroras. Drawings 1 and 6 are undetermined but could perhaps depict atmospheric optics or clouds. From a historical point of view, the datable aurora drawings are much older than previously known drawings of auroras and hence the earliest images of datable aurora drawn by the contemporary observer. Scientifically, their detailed descriptions of the morphology and motion indicate discrete auroras that are thought to be normally seen only in the polar region. Considering the relatively low geomagnetic latitude of Amida (45 degrees), they should have been strong geomagnetic storms, indicating strong solar activity around 771–773 CE. These results let us know the aurora activity under extreme space weather events as the values of these events are comparable with extreme cases with their formal Dst scales recorded. This autograph manuscript also provides information for comets comparable with known East Asian records for the same comets. These descriptions and drawings are of great importance as they provide primary information by the contemporary observers. Especially, Drawing 6 and the associated text are, to the best of our knowledge, the oldest evidence of the naked-eye observation of ion and dust tails, which are driven by radiation pressure and solar wind, respectively. Nevertheless, we do not dare to determine that they are definitely the very earliest drawings for auroras based on the contemporary observations, as we have still considerable volumes of unexamined manuscripts. Our finding suggests a possibility for other manuscripts to have earlier drawings. Our work is an opening for further astronomical surveys for ancient or medieval manuscripts.

**Acknowledgement**

We acknowledge support from the Center for the Promotion of Integrated Sciences (CPIS) of SOKENDAI as well as Kyoto University's Supporting Program for Interaction-based Initiative Team Studies "Integrated study on human in space" (PI: H. Isobe), the Interdisciplinary Research Idea contest 2014 by the Center of Promotion Interdisciplinary Education and Research, the "UCHUGAKU" project of the Unit of Synergetic Studies for Space, and the Exploratory Research Projects of the Research Institute of Sustainable Humanosphere, Kyoto University. This work was also supported by Grant-in-Aid from the Ministry of Education, Culture, Sports, Science and Technology of Japan, Grant Numbers JP15H05816 (PI: S. Yoden), JP26870111 (PI: Y. Mitsuma), and JP15H05815 (PI: Y. Miyoshi).

**Appendices**

**Appendix 1: Conversion of Seleucid Era**

In order to determine the observational dates, we need to convert the Seleucid Era (SE) to the Common Era (CE). SE starts at Dios 1 of 312 BCE (Macedonian Calendar) or Nisan 1 of 311 BCE (Babylonian Calendar). The starting point commemorates Seleucus I Nicator's reconquest of Babylon. This era was used in the Near East and accepted by the Syriac Orthodox Church (Parker & Dubberstein 1956; Harrak 1999).

According to Syriac Calendar, a year is consisted of the following 12 months: *tīšrī(n) qdēm* (October), *tīšrīn ḥrāyā* (November), *kānūn qadmāyā* (December), *kānūn trayyānā* (January), *šbāṭ* (February), *ādār* (March), *nīsān* (April), *êyār* (May), *ḥzīrān* (June), *tammūz* (July), *āb* (August), *êlūl* (September) (Thackston 1999). At the first council of Nicaea in 325 CE, the Syriac Orthodox Church, to which Zūqnīn monastery belonged, adopted the Julian Calendar to determine when to celebrate the Easter (Eusebius, Vita Const., III: 18-20; Percival 1900: p54; Edwards 2006: p553). Syrian Calendar was adjusted to Julian Calendar, although each year begins on the first day of *tīšrī(n) qdēm* corresponding to October 1. For the purpose of the calculation, the beginning of the year SE 1 was set on the 1st day in month *tīšrī(n) qdēm*, namely on 1 October 312 BCE, and the end of this year was set on the last (30th) day in month *êlul*, namely on 30 September 311 BCE.

Therefore we can convert the dates in the Seleucid Era and the Syrian Calendar into the dates in the Common Era and the Julian Calendar only considering their month.

**Appendix 2: Transliterations and English Translations**

Here we provide transliteration and English translation by the authors of the corresponding part of the Syriac chronicle Zūqnīn. The format of this chapter is consisted as follows: drawing number, Seleucid year, Common Era year, folio numbers of the original manuscript, page numbers in the critical editions by Chabot.

**Drawing 1:** SE811 (499/500 CE)

**Reference:** Vat.Sir.162, ff. 72v-73r, Chabot I: p263

**Transliteration:** *b-tešrī dēn ḥrāy ḥzaynan ātwātā tlāt ba-šmayyā ʿeddān pelgeh d-yawmā. ḥdā mennhēn ba-mṣaʿatāh da-šmayyā b-gabbā taymnāyā. d-dāmyā b-gawnāh l-qeštā d-hāwyā b-ʿnānē. wa-kpāpāh ḥāʾar lʿel: hānaw dēn kpāpāh ltaḥt w-rêšēh lʿel. wa-ḥdā men madnḥā: wa-ḥrētā tūb men maʿarbā.*

**Translation:** In *tešrī ḥrāy* (November) we saw three signs in the sky at midday. One of them in the middle of the sky, in southern side. It resembled in its color a bow that is in clouds. And its arc was looking upward, therefore it was its arc that was below and its ends were above. And one was in the east and another in the west.

**Reference:** Vat.Sir.162: f. 87v; Chabot II: p4

**Transliteration:** *w-tūb bāh b-šattā b-tešrī ḥrāy: etḥzīyēn ātwātā tlāt ba-šmayyā: b-pelgeh d-yawmā d-dāmyān l-qeštā d-metḥazyā b-ʿnānē. kad hū kpāphēn ltaḥt w-rêšhēn lʿel. ḥdā mennhēn b-gabbā taymnāyā b-meṣʿat šmayyā: wa-ḥrētā men madnḥā lāh. wa-ḥrētā men maʿarbā. w-hākannā hpīkān-way l-appay šmayyā.*

**Translation:** And again in this year, in *tešrī ḥrāy* (November), three signs were seen in the sky, at midday, (each of) them resembled a bow that is seen in clouds. Their arcs were equally below and their ends were above. One of them was in the southern side, in the middle of the sky, and another in its east. And another in the west. And thus they were turned toward the sky.

**Drawing 2:** 811 SE (499/500 CE)

**Reference:** Vat.Sir.162, ff. 72v-73r, Chabot I: p263

**Transliteration:** *w-tūb b-kānūn ḥrāy: ḥzaynan ātā ḥrētā men maʿarbā w-taymnā: bāh b-gōnyā d-dāmyā-wāt l-nayzkā: w-mennhon da-bnaynāšā āmrīn-waw ʿalēh d-maknaštā-y d-abdānā. w-mennhon tūb āmrīn-waw d-nayzkā-w d-ḥarbā.*

**Translation:** And again in *kānūn ḥrāy* (January), we saw another sign in the southwest. In the corner, it resembled a lance. And some people said about it that it was a besom of perdition. And others said that it was a lance of war.

**Drawing 3:** 811 SE (499/500 CE)

**Reference:** Vat.Sir.162: f. 87v; Chabot II 4

**Transliteration:** *w-tūb bāh b-šattā b-īrah kānūn etḥazyat ātā ḥrētā ba-šmayyā d-dāmyā-wāt l-nayzkā. b-gōnyā taymnāytā w-maʻrbāytā. d-mennhon da-bnaynāšā maknīštā d-abdānā qra'uh. w-mennhon tūb qra'uh nayzkā d-ḥarbā. ellā w-āp ršīm-ū lan šarbhēn men l'el.*

**Translation:** And again in this year, in *kānūn* (January), another sign, which resembled a lance, was seen in the sky, in the south-western side. Some people called it the besom of perdition. And others called it the lance of war. But their story is also written down by us above.

**Drawing 4:** 813 SE (501/502 CE)

**Reference:** Vat.Sir.162: f. 87v; Chabot II: p4

**Transliteration:** *šnat tmānemā w-tlātaʻsrē hwā zawʻā rabbā w-ethapkat pāṭelmaydā: w-ṣōr w-ṣīdōn. w-āp etʻaqrat w-neplat bēt šabbtā d-yūdāyē. w-beh b-lêlyā hānā d-zawʻā: da-hwā b-īraḥ āb b-ʻesrīn wa-trēn beh: b-maghay ʻrūbtā etḥazyat ātā b-gabbā garbyāyā ba-dmūt nūrā d-metgawzlā.*

**Translation:** The year eight hundred and thirteen, a big earthquake occurred and Ptolemais was overthrown, Tyre and Sidon too. And also a synagogue of Jews was uprooted and collapsed. And in the same night of the earthquake, which occurred in *āb* (August), the twenty second of the month, on the night preceding Friday, a sign in the shape of an inflamed fire was seen in the northern side.

**Reference:** Vat.Sir.162: f. 75r–v; Chabot I: p273–274

**Transliteration:** *hāšā dēn šmaʻ ʻal gūnḥē d-estʻar b-šattā hādē: w-ʻal w-ʻal(sic) ātā hāy d-etḥazyat b-yawmā haw d-beh estʻar. meṭṭul d-āp hādē hū tbaʻt b-īdan. b-yōm ʻesrīn wa-trēn b-īraḥ āb: d-šattā hādē b-maghay ʻrūbtā: nūrā saggītā etḥazyat lan kad metnabršā ba-pnītā d-garbyā kolleh lêlyā. w-hākannā sābrīn-wayn da-b-māmūlā d-nūrā ʻtīdā-wāt da-tkappar l-kollāh arʻā b-lêlyā haw. raḥmaw dēn d-māran naṭrun d-lā nekyānā.*

**Translation:** Now listen to the calamities which happened in this year, and to that sign which was seen on that day on which they happened, because you required this at our hand. On the twenty second day of *āb* (August) of this year, on the night preceding Friday, a great fire was seen by us, flaming in the northern region the whole night. And thus we thought that it would wipe the whole earth by a deluge of fire in that night. But the mercy of Our Lord kept us uninjured.

**Drawing 5:** 885 SE (573/574 CE)

**Reference**: Vat.Sir.162: f. 121v; Chabot II: p144–145

**Transliteration:** šnat tmānemā wa-tmānīn w-ḥammeš etḥzī nayzkā d-nūrā rabbā wa-dḥīlā ba-pnītā garbyāytā. b-yōm īraḥ êyār. kad šūrāyā hānaw dēn rêšeh taḥtāyā men kawkbā šāqel-wā leh. w-īt-wā leh metḥā rabbā. kad men qdīm l-appay pelgeh d-lêlyā sāleq-wā w-metḥzē. w-bātarkīn men ramšā. kad metḥzē-wā d-marken rêšeh l-madnḥā: w-kad qallīl qallīl traṣ-wā w-qā'em l'el trīṣā'īt: ba-dmūt rūmḥā rabbtā. w-tūb bātarkīn meṣṭlē. w-metrken l-appay ṣaprā la-pnītā ma'rbāytā. wa-hwā zabnā d-yarḥē trēn aw tlātā kad hākannā sāleq-wā w-metḥzē ammīnā'īt. kad bātarkīn hpak w-etḥzī tūb ba-pnītā taymnāytā. w-hākannā b-mawteh d-yūsṭīnānā malkā etḥappī w-lā tūb etḥzī. hālēn dēn lwāt rūšmē d-'ūhdānē: lwāt aylēn d-men bātarkīn aytīynan: d-kad qāreyn netbayyanūn b-ḥartā d-šūlāmhēn.

**Translation:** The year eight hundred and eighty and five, a big and dreadful lance of fire was seen in the northern region, on the days of the month êyār (May). At the beginning, it was its lower tip that proceeded from a star, and it extended at great length. At the beginning it used to rise and become visible at about midnight. And then from the evening. At that time it was visible that it bent its tip eastward. And then bit by bit it became straight and standing up straightly like a big spear. And then it leaned again and bent itself toward the west at about dawn. And it was for two or three months that it rose in this way and was seen constantly. Then it returned and was seen again in the southern region. And thus at the death of the emperor Justinian it veiled itself and no longer became visible. Now we brought these with the signs of remembrance to those who would be in the future, who would read and take notice of the result of their completion.

**Drawing 6:** 1071 SE (759/760 CE)

**Reference**: Vat.Sir.162: f. 136v; Chabot II: 217

**Transliteration:** šnat alpā w-šab'īn wa-ḥdā:

    b-īraḥ ādār: etḥazyat ātā ḥewwārtā ba-šmayyā: qdām šaprā. b-garbay madnaḥ. b-malwāšā haw d-metqrē emrā. l-garbyā men hālēn tlātā kawkbē: d-beh yattīr naṣṣīḥīn. w-dāmyā-wāt b-eskīmāh: l-maknaštā. kad tūb īt-wā beh b-emrā b-rêšeh: b-mūrā qadmāytā. b. men hālēn kawkbē ṭā'yā. qrāwnās w-arrês. ak da-l-taymnā qallīl. b-'esrīn wa-trēn beh b-yarḥā. w-kattrat hī ātā laylawwātā ḥammešta'sar. 'dammā l-nāgah 'êdā d-penṭêqosṭê. w-ḥad man rêšāh haw qaṭṭīnā. yattīr bahhūrā[3] kawkbā metḥzē-wā b-rêšeh: w-ṣālē-wā l-appay garbyā. haw dēn ḥrênā patyā w-yattīr

---

[3] *yattīr bahhūrā*, "duskier" is emended to *yattīr nāhūrā* by Chabot. H99 (p198) accepts this emendation and interpret the appearance of the end of the sign "more shining," contrasting with the appearance of another end, "darker (*yattīr 'ammūṭā*)." However, we can accept the original words, if

*'ammūṭā: ṣālē-wā l-appay taymnā. w-āzlā-wāt qallīl qallīl l-madnḥay garbay. eskīmāh dēn hāna-w:*

*nāgah dēn tlātā d-bātar penṭêqostê: etḥazyat tūb b-'eddān ramšā. men garbay ma'rābay. w-kattrat ramšē 'esrīn w-ḥammšā. w-āzlā-wāt qallīl qallīl l-taymnā.*

*w-tūb ebdat. w-haydēn hepkat etḥazyat b-ma'rābay taymnay. w-hākannā tammān kattrat yawmātā saggī'ē.*

*b-hānā zabnā sedqē saggī'ē hwaw b-'êdtā: men 'ellat rêšānūtā. kad 'ūmrē madnḥāyē 'bad l-yōḥannān paṭrīyārkā. kad mdīnātā da-gzīrtā lā šālmīn leh: w-āplā kollhon 'ūmrē. ma'rbāyē dēn w-mawṣlāyē l-gīwārgī šlem. w-men 'elltā hādē etdalḥat kollāh 'êdtā.*

**Translation:** The year one thousand and seventy and one.

In the month *ādār* (March), a white sign was seen in the sky, before early-morning, in the north-eastern side, in the zodiacal sign which is called Aries, to the north from those three stars, which are most shining in it. And it looks like in its shape, a besom. And it was still in Aries, at its head, in the first degree (of the sign), two (degrees) from those wandering stars, Kronos and Ares, which are slightly to the south, on the twenty-second of the month. And the sign remained for fifteen nights, until the dawn of the Pentecost feast. And one end of it was narrow and duskier, one star was seen in its tip, and it was turning to the north. And the other one, being wide and darker, was turning toward the south. And it (the sign) was going bit by bit to the northeast. This is its shape:

On the dusk of the third day after Pentecost, it appeared again in the evening in the northwest. And it remained twenty and five evenings. And it was going bit by bit to the south.

And it went astray again. Then it returned and was seen in the southwest, and in this way remained there many days.

In this time many sects were in the church, on account of the leadership. When the eastern monasteries made John Patriarch, neither the cities of Jazira nor all the monasteries approved him. People of the West and Mosul then approved George. And because of this, all the church was disturbed.

**Drawing 7:** 1080 SE (768/769 CE)

**Reference:** Vat.Sir.162: f. 146r; Chabot II: 257

**Transliteration:** *'al ātā d-ba-dmūtā d-maknīštā: d-etḥazyat ba-šmayyā.*

*ba-šnat alpā wa-tmānīn. b-īrah êyār etḥazyat ba-dmūt d-maknīštā: b-garbay madnaḥ. kad metḥazyā bāh b-gōnyā hādē. kad sālqā w-ḥeššōkā. ak haw d-ḥellā meddem d-baytā kānšā-wāt. 'am*

---

we think that neither of the ends were not bright, but their appearances were slightly different.

ṣaprā dēn metḥazyā d-marknā ṣōṣītāh l-appay arʿā. kad rādyā-wāt qallīl qallīl la-qdām(ē)h: ʿdammā d-ʿellat b-sarnā haw d-gīglā hāy d-ba-šmayyā. w-bāh etbalʿat w-ebdat. dmūtāh dēn hādē ītēh: hādē d-men lʿel ršīmā. ṭābāʾīt w-āp zādqāʾīt etqaryat maknīštā. meṭṭul d-aykannā d-ʿāʾel gārōpā āp maknīštā l-baytā w-gārpīn w-kānšīn leh: hākannā āp hādē knašteh l-ʿālmā. w-awbdat l-koll d-beh.

**Translation:** Concerning the sign, which was like a besom, and which was seen in the sky.

In the year one thousand and eighty, in the month of *êyār* (May), one like a besom was seen, in the northeast. When it was seen in this side, it was going up and dark, as if it was collecting a dust of a house. Then at dawn it was seen, bending its lock of hair toward the earth. It was moving forward bit by bit, until it entered the axis of the wheel in the sky, where it was swallowed up and went astray. Its shape is that which is drawn above. Rightly and duly it was called a besom, because as a spade and a besom enter a house and cleanse and sweep it, this also swept the world in this way, and brought everything in it to naught.

**Drawing 8:** 1083 SE (771/772 CE)

**Reference:** Vat.Sir.162: f. 150v; Chabot II: 275–277

**Transliteration:** ʿal qeštā d-māran d-etḥazyat b-hādē šattā kad hpīkā: w-ʿal šabṭā ḥewwārā d-etḥzī ba-mṣaʿtā da-šmayyā: da-trīṣ-wā lūqbal kpāpā dīlāh d-qeštā: ba-dmūt gêrā.

 tūb dēn āp bāh b-hādē šattā: b-āraḥ (sic) êyār etḥazyat qeštā hādē da-b-ʿnānē ammīnāʾīt metḥazyā: kad keptāh ltaḥt: wa-drāʿēh l-appay lʿel. w-metdamyā l-qeštā da-mtīḥā b-īdā d-gabrā la-qrābā. kad āp mḥawwyā lūḥāmā wa-gzāmā ʿal ʿāmōrēh d-arʿā. etḥazyat dēn b-yōm ḥadbšabbā qaddīšā ba-tlāt šāʿīn d-īmāmā: ak mā da-shed sābē myaqqrē hālēn da-ḥzaw qdāmayn. w-en nāš lā ṣābē la-mhaymānū l-hādē nebqor b-qeppallē qadmāyē w-meškaḥ akkūtāh d-hādē. kad āp sāhdān aylēn da-hway bātarkīn. tūb etḥzī ba-dmūt šabṭā ḥewwārā: kad mtīḥ men sawpāh maʿrbāyā da-šmayyā: w-ātē ʿdammā d-nīp rêšeh l-madnḥā ba-mṣʿatāh da-šmayyā. ʿabyāneh dēn ak ḥablā. w-lā (sic) tūb hānā l-saggīʾē etḥzī. meṭṭul d-yawmātā saggīʾē hwā kad sāleq. w-saggīʾē saggīʾātā emar ʿlaw. kad mennhon āmrīn: d-šabṭā ītaw d-rūgzā. ḥrānē āmrīn d-sārōqā ḥad men hālēn d-sālqīn ba-šmayyā ītaw. law meddem-ū. pārōšē dēn w-dāḥlay alāhā: kad ātā hādē ḥzaw: deḥltā rabbtā (e)tmlī. b-hāy d-īdaʿ d-ʿelltā da-ḥtāhē ītēh hādē: w-lūḥāmā d-rūgzā malyā. šāṭyē āplā ʿal bālhon aytī l-hādē: ḥakkīmā lam l-ṭawḥā ḥāʾar. saklā w-lā qdām ʿaynaw. ḥakkīmā lam ʿaynaw b-rēšeh. w-saklā b-ḥeššōkā rādē: trṣyh[4] dēn d-šabṭā hānā lūqbal mṣʿatāh d-qeštā da-mtīḥā ātē-wā.

---

[4] This word's pronunciation is uncertain (Payne Smith 1879–1901, col. 4511).

*w-ḥawwī sāʿārūteh w-lā saggī awḥar. wa-sʿar maʿbdānūtā hāy d-bāh eštaddar men alāhā. akbar emar nāš: d-law qeštā w-g(ê)rē īt leh l-alāhā. w-haw d-hālēn ṣābē l-mêmar: nešmaʿ la-mzammrāmrā mānā emar. nettrīm lam alāhā w-nešdē bhon gêrā men šelyā. w-netkarḥūn leššānayhon. w-net(da)ḥḥlūn koll d-ḥāzeyn lhon. w-ne(d)ḥlūn koll bnaynāšā. w-tūb emar. d-šaddar gêraw wa-bdar ennon. etbaddar bnaynāšā. hwaw gīyōrē b-koll atrawwān: ḥreb arʿātā. ṣaddī qūryas. ezal ʿammā atrā ʿal atrā.*

**Translation:** Concerning Our Lord's bow which was seen in this year and overturned, and concerning the white scepter which was seen in the middle of the sky, and which was directed toward the arc of the same bow, like an arrow.

Also in this year, in the month *êyār* (May), this bow was constantly seen in the clouds, its arc being below, and its shoulders being above. And it was like a bow which was stretched by the hand of a man for a battle, and also exhibiting menace and threat against the inhabitants of the earth. It was seen on a holy Sunday, at the third hour of the day, as these honorable old men who saw it testified before us. If someone does not want to believe this matter, he should seek in the previous chapters and he will find an equivalent of this. And those which appeared afterwards also testify. Then something like a white scepter was seen, extended from the western end of the sky, and it came to the middle of the sky, until its head turned toward the east. Then its swelling was like a rope. And this was not (sic) seen by many people, because it ascended for many days. Many people said many things about it. While some people said that it was the scepter of rage, others said that it was one of clouds those which ascended to the sky, it was nothing. Then the people with discernment and God-fearing ones, when they saw this sign, they were filled with great fear, for they knew that it was an evidence of sins and was full of menace of rage. Stupid people did not mind about it. "A wise man looks far away, but a stupid man not even in front of his eyes." "A wise man has his eyes in his head, but a stupid man proceeds in darkness." Then this scepter directed toward the center of the extended bow. And it showed its effect and did not delay much. And it did the operation for which it had been sent by God. Perhaps someone said that God has neither a bow nor arrows. Let the one who wants to say these (words) listen to what the Psalmist said: "God will go up and suddenly shoot arrow at them. Then their tongues will languish. All who see them will be scared, and all men will fear (Ps 64:8-9)." He also said: "He sent out his arrows and scattered them (Ps 18:15)." People scattered. They were aliens in all the regions. Lands were ruined, villages became deserted. People went from one region to another.

**Drawing 9:** 1083 SE (771/772 CE)

**Reference:** Vat.Sir.162: f. 150v; Chabot II: 277

**Transliteration:** 'al ātā ḥrētā d-etḥazyat tūb ba-pnītā garbyāytā bāh b-šattā.

ātā dēn ḥrētā etḥazyat b-sawpā garbyā(yā). kad hū ḥezwāh msahhed-wā 'al lūḥāmeh wa-gzāmeh d-alāhā da-'layn. etḥazyat dēn b-yawmātā da-ḥṣādā: kad lābkā-wāt l-kolleh gabbā garbyāyā men qarnā mad(n)ḥāytā wa-'dammā l-qarnā ma'rbāytā. dmūtāh dēn hākannā ītēh: ḥad šabṭā sūmāqā: w-ḥad yūrāqā. w-ḥad ūkāmā. w-ḥad kūrkmānā. sālqā-wāt dēn men ltaḥt wa-'dammā l'el. kad ḥad šabṭā dā'ek: wa-ḥrênā sāleq[5]. w-kad nāš ḥā'ar bāh mešṭaḥlpā-wāt l-šab'īn šūḥlāpē. wa-l-pārōšē ātā da-gzāmā mawd'ā-wāt. w-'al hādē saggī'ē saggī('ā)tā emar. kad āmrīn da-dmā mawd'ā. ḥrānē ḥrānyātā emar. ellā mannū yāda' 'bādaw d-māryā. ettel lam ātwātā ba-šmayyā w-gabbarwātā 'al ar'ā.

**Translation:** Concerning another sign which was seen in the northern region also in this year.

Another sign was seen in the northern end, its appearance testifying to the menace and threat of God against us. It was seen at harvest time, occupying the entire northern side from the eastern corner to the western corner. Its form was as follows: a blood-red scepter, a green one, a black one, and a saffron-colored one. It was going from below to above. When one scepter was extinguished, another one went up. And when someone was looking at it, it was changed into seventy shapes. To the people with discernment it showed the sign of threat. And many people said many things about it. While some said it announced bloodshed, others said other things. But who knows the deeds of the Lord? "I will give signs in the heaven and miracles on the earth (Acts 2:19)."

**Drawing 10:** 1084 SE (772/773 CE)

**Reference:** Vat.Sir.162: f. 155v; Chabot II: 301–302

**Transliteration:** 'al ātā hāy qadmāytā d-etḥazyat-wāt ba-pnītā garbyāytā. etḥazyat āp b-hādē šattā.

alāhā men qdīm ba-nbīyē mallel-wā lwāt haw bīšā (m)marmrānā. l-ḥartā dēn mallel b-yad breh ḥabbībā: 'am kolleh yaldeh d-ādām. w-ḥāšā lan dēn bnayyā margzānē d-īt lan mellayhon da-nbīyē: w-šūdā'ē hālēn d-yab pārōqā l-'êdteh w-kār(ō)zūtā da-šlīḥē: w-'beyn lebbayn wa-'mīṣān 'aynayn w-šī'ān ednayn: d-lā neḥzē b-'aynayn w-nešma' b-ednayn: w-nestakkal b-lebba(y)n l-mellaw ḥayyātā d-pārōqan: wa-ntūb men bīšātan w-nêḥē. etḥzī lan ātwātā ba-šmayyā da-mḥawwyān gzāmē:

---

[5] This active participle, *sāleq*, is translated "appearing" by H99. However, the feminine of the same participle, *sālqā*, is used with the words "from below to above (*men ltaḥt wa-'dammā l'el*)" in the previous sentence for indicating the upward movement or upright extension of the entire "sign (*ātā*)." Since the verb *sleq* itself can be translated "to go up, ascend" (Payne Smith & Payne Smith 1903, 379), the masculine participle indicates an upward movement or upright extension of "scepter (*šabṭā*)" rather than its simple appearance.

*ak da-lwāt ʿammā lā met(ṭ)pīsānā. d-hū ḥezūhēn msahhed-wā l-pārōšē ʿal sōgā d-bīšātan: w-ʿal rūgzā haw da-ʿlayn lḥīm-wā men lwāt kênūtā. ātā hāy d-men qdām šattā etḥazyat-wāt ba-pnītā garbyā(y)tā: etḥazyat-wāt āp b-hādē šattā. b-āraḥ (sic) ḥzīrān. b-yōm ʿrūbtā. kad āp hālēn tlāt šnīn: d-etḥazyat ḥdā bātar ḥdā: b-yōm ʿrūbtā metḥazyā-wāt. kad āp māṭḥā-wāt napšeh (sic) men gabbā madnḥāyā ʿdammā l-gabbā maʿrbāyā. w-kad qāʾem nāš la-mḥār bāh metḥallpā-wāt l-šūḥlāpē saggīʾē. aykannā d-mā da-dʿek zallīqā sūmāqā: sāleq yūrāqā. w-mā da-dʿek yūrāqā sāleq kūrkmānā. w-mā d-hānā dʿek sāleq ūkāmā*[6]. *kad mawdʿā d-law balḥōd ḥad ūlṣānā sāblā arʿā: ellā w-da-mšaḥlpīn men ḥdādē. ak mā da-gdaš lan ba-ʿbādā: dmūtā dēn d-ātā hādē hāy da-ršīmā men lʿel.*

**Translation:** Concerning the previous sign which was seen in the northern region. It was seen also in this year.

God formerly spoke to that rebellious evil through the prophets. And later he spoke to every child of Adam through his beloved son. And now to us, the irritating sons, who have the words of the prophets, and those promises which the Saviour has given to his Church, and the preaching of the Apostles. But our hearts are hardened, and our eyes are closed, and our ears are stopped, so that we may neither see with our eyes nor listen with our ears, nor understand with our minds the living words of our Saviour nor repent of our wickedness nor live. Signs were seen by us in the sky, showing threats, like which are against disobedient people, and their appearance testified to the people with discernment about the multitude of our wickedness, and about the rage which is menacing us with justice. The sign which was seen a year ago in the northern region, was also seen in this year, in the month *ḥzīrān* (June), on a Friday. And equally in these three years, in which one (sign) after another was seen, each appeared on a Friday. And each equally stretched from eastern side to western side. And when someone rose up to watch it, it was changed to many different shapes in this way: when a blood-red ray was extinguished, a green one went up, and when the green one was extinguished, a saffron-colored one went up, and when this one was extinguished, a black one went up. So it announced that the land would not suffer only one affliction, but various ones from others, as it really happened to us. And it is the shape of this sign that is drawn above.

---

[6] The participle *sāleq* is used three times here and every time translated "(would) appear" by H99. However, it is probably used to indicate upward movements or upright extension of colorful "rays (*zallīqē*)", because the same participle in Record 4b shows such movements or extension of colorful "scepters (*šabṭē*)" (see note 3).

Table 1: Summary of drawings in MS Vat.Sir.162.

| Drawing | folio number | year in CE (SE), month | interpretation | Note |
|---|---|---|---|---|
| 1 | 72v-73r | 499/500 (811) | Halo | |
| 2 | 72v-73r | 499/500 (811) | Comet | |
| 3 | 87v | 499/500 (811) | Comet | |
| 4 | 87v | 501/502 (813) | Aurora | |
| 5 | 121v | 573/574 (885) | Comet C574/G1 | With caption |
| 6 | 136v | 759/760 (1071) | Halley's Comet | Two tails mentioned |
| 7 | 146f | 769 (1080) | Comet C770/K1 | Year likely 770. |
| 8 | 150v | 772 (1083) | Cloud? | |
| 9 | 150v | 771/772 (1083) | Aurora | Observed "at harvest" |
| 10 | 155v | 773 (1084) | Aurora | Friday |